\documentstyle[twoside,fleqn,epsf,espcrc2]{article}

\def\be		{\begin{equation} }
\def\ee		{\end{equation} }
\def\bea		{\begin{eqnarray} }
\def\eea		{\end{eqnarray} }

\newlength{\figsize}
\newlength{\figoffset}
\newlength{\figbackup}
\newlength{\figendsp}

\title{Monopoles and Vortices in the SU(2) Positive Plaquette Model}

\author{John D. Stack
	\address{Dept. of Physics, University of Illinois 
	at Urbana--Champaign, 1110 W. Green Street, 
      	Urbana, IL 61801, USA.\\}
\thanks{Talk presented by J. Stack}
	and William Tucker$^a$}%

\begin{document}

\begin{abstract}
\noindent
We study the heavy quark potential in the SU(2) positive plaquette model
using monopoles in the maximum abelian gauge, and vortices.  Monopoles
give a quantitative description of the string tension.  Vortices approximately
reproduce the entire heavy quark potential.  
\end{abstract}

\maketitle

After many years of work, monopoles and vortices have emerged as the main
possibilities for explaining confinement.  Most research has been carried
out for the Wilson action in SU(2) lattice gauge theory.  Here we study
the case of the positive plaquette model (PPM) which weights configurations
in the same way as the Wilson action for positive plaquettes, but suppresses
negative plaquettes entirely \cite{fingberg}.  A negative plaquette 
represents a
huge field strength of order $1/a^{2}$,  an obvious lattice artifact.
Since negative plaquettes are suppressed in the PPM, the PPM should give a
clearer view of continuum physics.  Our goal here is to see how closely
monopoles and vortices describe the long range confining physics 
in the PPM.  We work on a $16^{4}$ lattice at couplings $\beta_{PPM}
= 1.790, \, 1.840, \, $ and 1.886.  These couplings were chosen by using
the accurately determined deconfining temperature $T_{c}$ 
for the PPM \cite{fingberg}. In particular, $T_{c}=1/8a$ corresponds
to $\beta_{PPM}=1.886(6)$.  To find a value of $\beta_{PPM}$ appropriate
for use on a $16^{4}$ lattice, we assumed the ratio $T_{c}/\sqrt{\sigma}$
(where $\sigma$ is the string tension) is universal.  Using
$T_{c}/\sqrt{\sigma}=0.69(2)$, which is the known value for the Wilson 
action \cite{fingberg2}, we find that $\xi =1/\sqrt{\sigma} \sim 5.5a$
for $\beta_{PPM}=1.886(6)$.  Thus $\beta_{PPM}=1.886(6)$ has a similar
value of correlation length $\xi$ to $\beta_{W} =2.50$. We generated 
500 well-separated PPM configurations
at this $\beta_{PPM}$ and the two neighboring ones listed in Table~1.

\begin{figure}[t]

\leavevmode
\begin{center}

\hbox{%
\hspace{\figoffset}
\epsfxsize = \figsize
\epsffile{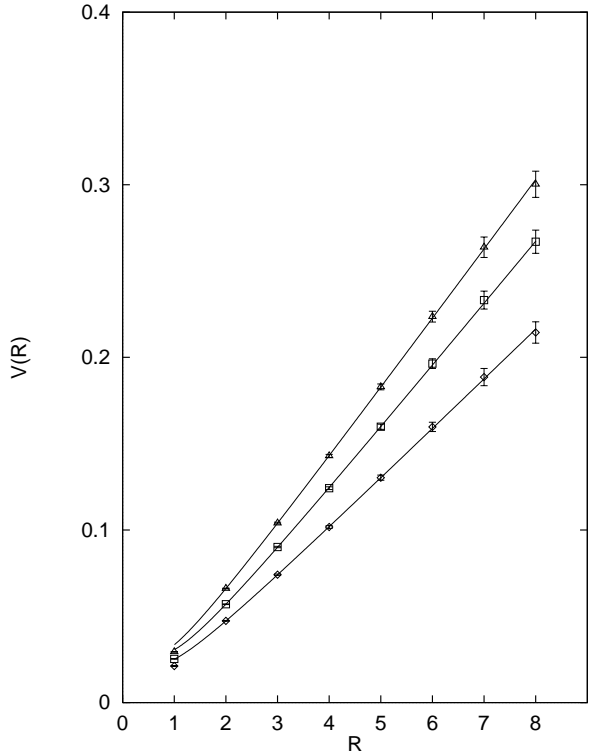}
}

\vspace{\figbackup}
\end{center}

\caption{Monopole potentials in PPM at
$\beta_{PPM}=1.790$ (triangles), 1.840 (squares),
and 1.886 (diamonds)}
\label{fig_one}

\vspace{\figendsp}
\end{figure}

For the case of monopoles,  the configurations were 
put into the
maximum abelian gauge and the magnetic current extracted.  From the magnetic
current, the magnetic vector potential, $A^{m}_\mu(x)$ was constructed,
\begin{equation} A^{m}_\mu(x)=
\sum_{y}v(x-y)\bar{m}_{\mu}(y).
\label{eqnamag}
\end{equation}
Then the monopole contribution to a Wilson loop, $W_{mon}$, was calculated,
\begin{equation}
W_{mon} =
                \left<\exp\left(\frac{i2\pi}{2}\sum_{x}
D_{\mu\nu}(x)
                   F^{*}_{\mu\nu}(x)\right)\right>_m ,
\label{eqnmon}
\end{equation}
where $F^{*}_{\mu\nu}(x)$ is the dual of the field strength constructed from
$A^{m}_\mu(x)$.  The resulting heavy quark potentials are shown in 
Figure~\ref{fig_one} , and the string tensions {\it vs.} those for full PPM
SU(2) are shown in Table~1.
\begin{table}
\begin{center}
\begin{tabular}{||c|c|c||} \hline
$\beta_{PPM}$ & $\sigma_{SU(2)}$ & $\sigma_{mon}$\\ \hline
1.790 & 0.0426(6) & 0.043(2) \\ \hline
1.840 & 0.036(1) & 0.036(1) \\ \hline
1.886 & 0.029(1) & 0.028(2) \\ \hline
\end{tabular}
\end{center}
\begin{center}
{\bf Table 1}
\end{center}
\end{table}

In our previous work on the PPM \cite{stack_neiman}, the full SU(2) 
string tension was 
evaluated slightly incorrectly, since we used a multi-hit program
to calculate Wilson loops which updated the links along the sides of
the loop using the Wilson action instead of the PPM action.  Correcting
this made only a minor change at $\beta_{PPM}=1.790$, shown in Table~1.
The other two $\beta_{PPM}$ values are the old numbers.  The important
thing to note in Table~1 is the excellent agreement between monopole
and full SU(2) string tensions.

It is of interest to ask what is the difference between the monopole
description of the heavy quark potential for the PPM and the Wilson action.
At $\beta_{PPM}=1.840$, we have $\sigma_{mon} = 0.036(1)$, very close
to $\sigma_{mon}= 0.034(1)$ at $\beta_{W}=2.50$ for the Wilson action 
\cite{jssnrw}.
However, for the number of links with magnetic current, we find
2936(25) for the PPM, whereas the Wilson action has  3565(22), so
there are $\sim 600$ more links carrying magnetic current for the
Wilson action. The origin of this difference becomes clear when the
magnetic current is resolved into loops.  It is known that only large
loops contribute to the string tension.  When loops with less than 50 links
of magnetic current are dropped, the string tension retains the same value
for both actions.  Comparing the number of magnetic current
links for loops larger than
50 links, we now find quite comparable results, 
1520(20) for the PPM, and 1573(20) for the Wilson action.
The excess of $\sim 600$ links found for the Wilson action is then mainly in
small loops which do not affect the string tension.  Suppression of negative
plaquettes suppresses these small loops, or in other words,
suppression of an obvious lattice
artifact gives a magnetic current more concentrated in the large loops known
to be relevant for continuum physics.

Turning now to vortices, there is a long history of work initiated by Mack and
collaborators, \cite{mack4}, and Tomboulis and collaborators \cite{tomb_kov},
which seeks to explain confinement via vortices associated with the center
of the gauge group, Z(2), for the present case of an $SU(2)$ gauge group.
In SU(2), vortices can be classified as `thin' (composed of negative
plaquettes), `hybrid' (composed partly of negative plaquettes), and
`thick' (composed entirely of positive plaquettes) \cite{tomb_kov}.
In the PPM, there can only be thick vortices.  In all cases the 
Wilson loop is a vortex counter, the vortex contribution to a Wilson
loop being
\begin{equation}
W_{vort} = \langle (-1)^{n} \rangle,
\end{equation}
where $n$ is the number of vortices piercing the Wilson loop.  Vortices
are supposed to control the long range physics, and therefore $W_{vort}$
should produce the same string tension as full SU(2).  This has been
checked for the Wilson action \cite{tomb_kov}, but the PPM provides a more
stringent test, since it only allows thick vortices.  Calculation
of $W_{vort}$ is particularly simple; one just replaces the loop by its
sign.  To enhance statistics, we smeared the ends of the loops.  The overall
appearance of the potential derived from $W_{vort}$ is rather similar to
the full SU(2) potential as seen in Figure~2.  
 
\begin{figure}[htb]

\leavevmode
\begin{center}

\hbox{%
\hspace{2em}
\epsfxsize = 2.75in
\epsffile{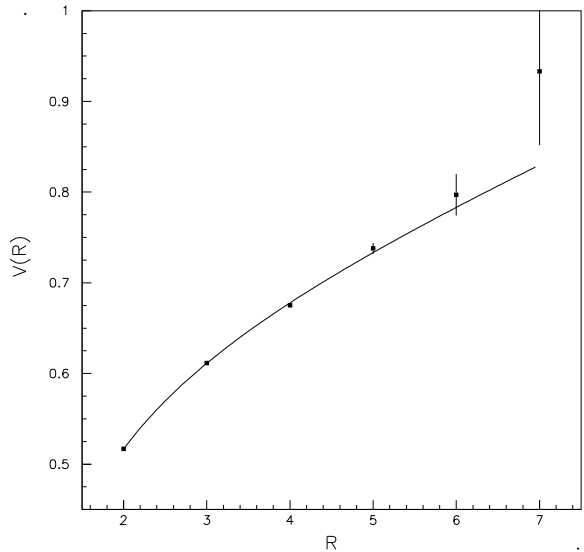}
}

\vspace{-15ex}
\end{center}

\caption{The potential from vortices
at $\beta_{PPM}=1.790$ }
\label{fig_two}

\end{figure}
The noise level is comparable, and the
potential has the same shape even in the Coulomb region at small R.
This is in contrast to the monopole potential.  Monopoles are quiet; no
smearing or multi-hitting is needed to see the potential out to R=8a, and
the monopole potential clearly differs from full SU(2) in
the Coulomb region, being basically linear at all values of R.  We have
extracted PPM results for the vortex string tension $\sigma_{vort}$ at only one
coupling so far, which we show in Table~2. 
As can be seen there, the
vortex string tension is somewhat low compared to monopoles and full SU(2).
A possible explanation is that
thin and hybrid vortices are still alive and
contributing to the string tension for the Wilson action, where it was
found that $\sigma_{vort}=\sigma_{SU(2)}$ \cite{tomb_kov}.  
Since the PPM has only
thick vortices, this would naturally give it a lower string tension.  Of
course, as the value of $\beta_{PPM}$ is increased, thin and hybrid vortices
are heavily suppressed, and only thick vortices remain. In this limit
presumably, we will have $\sigma_{vort}=\sigma_{SU(2)}$ for the PPM as
well as the Wilson action.

Assuming the small discrepancy for $\sigma_{vort}$ goes away with 
increasing $\beta_{PPM}$, one then has two viable descriptions of
the long range confining physics, one from monopoles and the other
from vortices.  In the future, we plan to work in two directions.
The first is to make the vortex picture more concrete by starting a
program of vortex location.  This we plan to do {\it without} gauge-fixing
by using small Wilson loops {\it e.g.} $2 \times 2,\; 3\times 3,\;
4\times 4\; etc.$ as vortex detectors.  
By this means, the physical picture of a large Wilson loop as pierced by
many vortices can be tested.  Our second line of investigation is to
try to see what connection there is between the monopole and vortex
descriptions, in particular to investigate the magnetic current
distribution near vortices.

\begin{table}
\begin{center}
\begin{tabular}{||c|c|c|c||} \hline
$\beta_{PPM}$ & $\sigma_{SU(2)}$ & $\sigma_{mon}$ & $\sigma_{vort}$\\ \hline
1.790 & 0.0426(6) & 0.043(2) & 0.036(2) \\ \hline
\end{tabular}
\end{center}
\begin{center}
{\bf Table 2}
\vspace{-7ex}
\end{center}
\end{table}

\noindent
This work was supported in part by the
National Science Foundation under Grant No. NSF PHY 94-12556.
%
%
%
%
%



\begin{thebibliography}{99}

\bibitem{fingberg} J.\ Fingberg, U.\ M.\ Heller, and 
V.\ Mitryishkin,  Nucl. Phys. 435 (1995) 311.

\bibitem{fingberg2} J.\ Fingberg, U.\  Heller, and 
F.\ Karsch,  Nucl. Phys. 392 (1993) 493.

\bibitem{stack_neiman}
J. D. Stack and S. D. Neiman,
Physics Letters B 377 (1996) 113.

\bibitem{jssnrw} J.\ D.\ Stack, S.\ D.\ Neiman, and
R.\ J.\ Wensley,  Phys. Rev. D  50 (1994) 3399.

\bibitem{mack4}G.\ Mack and V.\ B.\ Petkova
Ann. Phys. 123 (1979) 442.


\bibitem{tomb_kov} E.\ T.\ Tomboulis and T.\ G.\ Kov\`{a}cs,
Phys. Rev. D 57 (1998) 4054.


\end{thebibliography}
\end{document}